\documentstyle[editedvolume]{crckapb} 
\input epsf

\def\eg{e.g.,\ }

\def\etal{\it et~al.\rm\ }
\def\sun{\ifmmode_{\mathord\odot}\else$_{\mathord\odot}$\fi}
\def\gtsima{$\; \buildrel > \over \sim \;$}

\begin{opening}
\title{Dynamics of Mergers}

\author{Chris Mihos}
\institute{Department of Astronomy,
           Case Western Reserve University}

\end{opening}

\runningtitle{Dynamics of Mergers}

\begin{document}


\noindent The title ``Dynamics of Mergers'' seems broad enough to cover a daunting range of topics.
To make things more manageable, I will focus the discussion here on the type of mergers
which are largely thought to give rise to the subject of this workshop -- the ultraluminous
infrared galaxies (ULIRGs). Because of their morphologies and gas contents, ULIRGs are believed
to arise from mergers of two comparable mass, gas-rich spiral galaxies -- this will define the
``merger'' part of the title. The ``dynamics'' involved will be largely the gravitational 
dynamics of merging, and the dynamics of gas inflows which fuel the central activity in
ULIRGs, be it starburst or AGN. 

\section{The Life of a Merger}

To begin a discussion of merger dynamics, it is perhaps best to describe 
the different dynamical phases of the merging process. Figure 1 shows a ``typical'' merger (described in detail in Mihos \& Hernquist 
1996 [MH96]) involving two equal mass disk galaxies colliding on a parabolic orbit with perigalactic 
separation of $2.5h$, where $h$ is the exponential disk scale length. One disk is exactly prograde, 
while the other is inclined 71$^\circ$ to the orbital plane. Both disks are embedded in truncated 
isothermal dark halos with mass 5.8 times the disk mass. The half-mass rotation period is $t_{rot}$.

\begin{itemize}

\item {\sl Pre-collision} $\Delta t \sim 0.01({r_{init}\over h})^{3/2}t_{rot}$\hfil\break
As the galaxies fall in towards each other for the first time, they move on simple
parabolic orbits until they are close enough that they have entered each others' dark halos, and
the gravitational force becomes non-Keplerian. During this infall, the galaxies hardly respond
to one another at all, save for their orbital motion.
\smallskip

\item {\sl Impact!} $\Delta t \sim 0.3({r_p\over h})^{3/2}t_{rot}$\hfil\break
As the galaxies reach perigalacticon, they feel the strong tidal force from one
another. The galaxies become strongly distorted, and the tidal tails are (appropriately!)
launched from their back sides. Strong shocks are driven in the galaxies' ISM 
due to tidal caustics in the disks as well as direct hydrodynamic compression of the colliding 
ISM.
\smallskip

\item {\sl (Self-) Gravitational Response} $\Delta t \sim t_{rot}$\hfil\break
As the galaxies separate from their initial collision, the disk self-gravity can
amplify the tidal distortions into a strong $m=2$ spiral or bar pattern. This
self-gravitation response is strongly coupled to the internal structure of the  
galaxies as well as their orbital motion, resulting in a variety of dynamical
responses (see \S 3).
\smallskip

\item {\sl ``Hanging Out''} $\Delta t \sim $?? -- long? short?\hfil\break
Having plowed through the densest parts of one another's dark halos, the galaxies experience 
strong dynamical friction, causing the orbit to decay. The galaxies linger at apogalacticon
for a significant time (several to many rotation periods) before falling back together
and merging. The timescale here is crucially dependent on the distribution of dark
matter at large radius, resulting in significant uncertainties in the duration of
this phase (see \S 5).
\smallskip

\item {\sl Merging} $\Delta t \sim {\rm a\ few\ } t_{rot}$\hfil\break
Once the galaxies fall back together, they typically dance around each other once or twice
more on a short-period, decaying orbit before coalescing into a single remnant. During
this period, gravitational torques and hydrodynamic forces are strong, resulting in
strong gaseous inflow and rapid violent relaxation.
\smallskip

\item {\sl Relaxation} $\Delta t \sim {\rm\ a\ few}\ t_{rot}(R)$\hfil\break
Once the galaxies merge, a general rule of thumb is that violent relaxation and/or dynamical
mixing occurs on a few rotation periods {\it at the radius in question}. In the inner
regions, the remnant will be relaxed in only $\sim 10^8$ yr; in the outer
portions mixing may take $> 10^9$ yr.

\end{itemize}

\begin{figure}
\centerline{\epsfbox{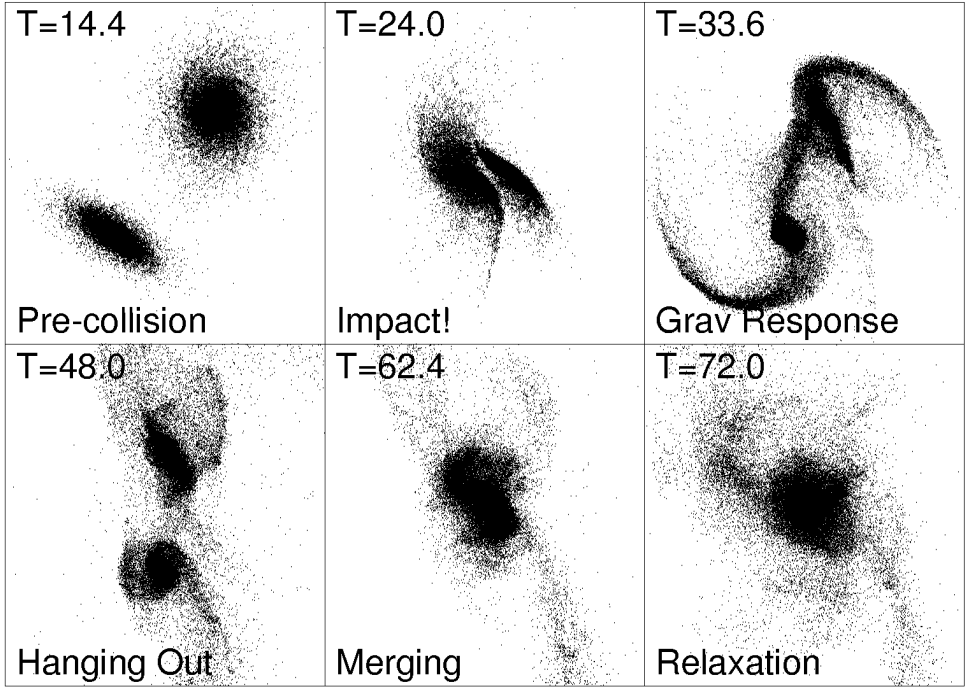}}
\caption{Merger model from MH96, illustrating the different dynamical phases.}
\end{figure}

\section{Where are the ULIRGs?}

Given this range of dynamical states, it is useful to ask which state preferentially hosts
ULIRG galaxies. The fact that they are predominantly close pairs or single, disturbed systems
argues that late stage systems dominate, but can we be more quantitative? Such insight can
come from an analysis of the projected separations of ULIRGs (\eg Murphy \etal 1996). If
we know the orbital evolution of binary galaxy pairs, we can statistically reconstruct the 
distribution of dynamical phases from the observed projected separations. In essence, this 
exercise will reveal how ULIRG activity samples the general merging population.

This selection function can be determined (in an admittedly model-dependent fashion) using N-body 
simulations of merging galaxies. A suite of merger models is calculated, focusing on the close ($r_{peri} =$ 2, 4, 6, 8, 
and 10 disk scale lengths), equal-mass mergers thought to give rise to ULIRGs. Given the orbital 
evolution of these models, we ``observe'' the model pairs randomly in projection, and 
weighted by $r_{peri}^2$ (geometric weighting of orbits). Because the merging timescale
differs drastically in mergers of different impact parameter, we define a {\it relative}
timescale as $t_{rel}=t/t_{merge}$ in which initial impact occurs at $t_{rel}=0$ and final
merging occurs at $t_{rel}=1$. 

If we ``observe'' the merger models completely randomly in time -- in essence assuming the the ULIRG
selection function is constant over dynamical stage -- we construct the histogram of projected
separation $\Delta R$ shown in Figure 2a. Because binary galaxies spend most of their
orbital lifetimes at apogalacticon, and because distant encounters are assumed to be more common
than close ones, $N(\Delta R)$ shows a strong peak at extremely wide separations, 
$\Delta R>$ 30 kpc. As this is far from the observed situation, a flat selection function 
(Fig 2c) is clearly unrealistic. However, from this histogram, we can do a Monte Carlo rejection of
observations in each bin until we match the true observed $N(\Delta R)$
histogram (Fig 2b). At this point, we can determine from the models the distribution of
dynamical ages of the surviving observations (Fig 2d). 
From this distribution, we see that ULIRGs must come predominantly from mergers in the final 20\%
of their merging history -- in other words, the final merging phase. A small fraction of objects
may come from objects near their initial collision. This selection function argues that galaxies
are somehow stable against the onset of ULIRG activity over most of the merging history, even
though they respond dynamically at a much faster pace. What, then, causes this disconnect between
dynamical response and ULIRG activity?

\begin{figure}
\centerline{\epsfbox{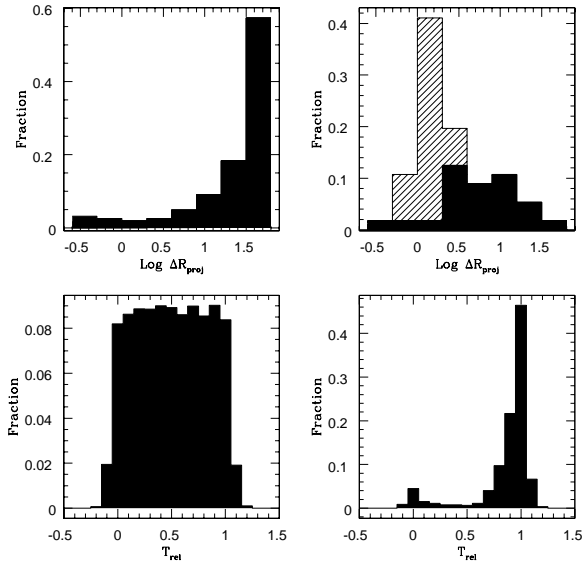}}
\caption{Monte Carlo ``observations'' of merger models. If orbital time is sampled uniformly
(bottom left), the distribution of projected separation is strongly weighted to distant pairs
(upper left). To match the observed distribution of ULIRG separations (upper right, from Murphy \etal
(1995); hatched regions represent upper limits), ULIRGs must be strongly biased towards late stage mergers (bottom right).
See text for details.}
\end{figure}

\section{Is there a Dynamical Trigger?}

{\it Whatever} powers the extreme luminosity of ULIRGs, the requisite is sufficient fuel in the
form of interstellar gas. From a dynamical point of view, the link between ULIRG activity and the merging 
process must lie in the detailed dynamics which drive nuclear gas inflows in mergers. Is there a distinct dynamical
trigger which begins this inflow and resultant ULIRG activity, or are there several paths to the formation
of ULIRGs?

\subsection{Theoretical Expectations}

Much of our understanding of the dynamical triggering of inflows in galaxies comes from
N-body simulations. These simulations have generally shown that gaseous inflows in galaxies
arise largely in response to the growth of $m=2$ instabilities in disks -- spiral arms or,
more strongly, bars (Noguchi 1988; Barnes \& Hernquist 1991; MH96; BH96). As such the question 
of inflow triggers becomes one of bar instability
in disks. What kind of encounters drive bars? When in the merging sequence do they form?

A variety of simulations have revealed a variety of answers. In disks which are susceptible
to global instabilities, strong bars form shortly after the initial collision.
In these situations, rapid inflow occurs within a few
disk rotation periods, providing the fuel for early starburst or AGN activity well before the
galaxies merge (MH96, BH96). If these types of galaxies were the dominant sources for ULIRGs, ULIRG
samples should contain many more wide pairs than are actually observed. Disk stability, therefore,
may be one criterion for forming ULIRGs. 

That stability may come from the presence of a massive central bulge or a low ratio of 
disk-to-dark matter in the inner disk. Simulations by Mihos \& Hernquist (1994, MH96) show
that a bulge component can stabilize the disk against bar formation, holding off inflow until the
galaxies ultimately merge. At this point the strong gravitational torques and gasdynamical
shocks overwhelm any stability offered by the bulges, and the gas is rapidly driven inwards
on a dynamical timescale, presumably fueling a starburst or active nucleus. In interesting
contrast to these models are those of Barnes \& Hernquist (1996, BH96) which employed a similar
3:1 disk-to-bulge ratio in their model galaxies, but with a much lower density bulge. In these
models, the bulges were unable to stabilize the disks, and early inflow again occurred. Clearly
it is thus more than the mere presence of bulges that stabilize disks -- the bulges must
be sufficiently concentrated that the dominate the mass distribution (and thus the rotation
curve) in the inner disk. Alternatively, a high fraction of dark-to-disk mass in the inner
portion of the galaxies may also stabilize the disks; such is probably the case in low
surface brightness disk galaxies (Mihos \etal 1997).

Aside from internal dynamics, the orbital dynamics also play a role in triggering inflow
and activity.  BH96 also show how orbital geometry influences the
inflow and activity. For galaxies with a modest amount of disk stability, a prograde encounter
will be sufficient to drive bar instabilities, while a retrograde encounter will not.
In these cases, retrograde disks will survive the initial impact relatively undamaged,
and not experience any strong activity until the galaxies ultimately merge. In the
extreme situation of very strong or very weak stability, however, internal stability
effects tend to win out over orbital effects (MH96).

More recently, simulations have shown that triggering of activity may not be solely tied to
the physics of inflows. Instead, the fueling of activity may be moderated by 
starburst energy, which can render the gas incapable of forming stars. Simulations
by Gerittsen (1998) indicate that a starburst can heat a significant fraction of the inflowing
gas to a few million degrees; the onset of star formation in this gas must 
await radiative cooling, resulting in milder but longer-lived starbursts compared to those
of MH96.
With the current uncertainties in modeling the physics of star formation and starburst feedback,
this result does not bode well for the detailed predictive power of {\it any} current starburst merger model.

\subsection{Observational Constraints}

While the dynamical models can guide our expectations, the variances due to effects such
as galactic structure, orbital geometry, and starburst physics make it hard to isolate
any single effect as a dominant trigger. Can we instead turn to the observational data to 
complement these models? 
Because of the rapid decoupling of the nuclear gas from the global kinematics, studies of
nuclear kinematics may give an improper account of the dynamical history of the encounter. 
Instead, global kinematics have a better ``memory'' of the initial conditions and evolution
of the collision.

To study these global kinematics, Mihos \& Bothun (1998) recently examined the two dimensional
H$\alpha$ velocity fields of four southern ULIRGs. These galaxies were chosen to display
extended tidal features, and thus biased the sample towards largely prograde systems. Nonetheless,
the four systems showed a wide range of kinematic structure, with no distinct commonality.
One (IRAS 14348-1447) showed short tidal features, extended H$\alpha$ emission, and fairly
quiescent disk kinematics, suggesting the system is a young interaction. The second (IRAS 19254-7245,
the Superantennae) possessed extremely long tidal features, more concentrated H$\alpha$ emission,
and evidence for outflowing winds -- clearly a more advanced interaction, although the pair is still
separated by $\sim$ 10 kpc and simple disk kinematics still dominate the overall velocity field. 
IRAS 23128-4250 is the most distorted
of the four, with two nuclei separated by $\sim$ 4 kpc, several distinct overlapping kinematic
components, and a 90$^{\rm o}$ slew in the angular momentum vector of the system from the
nuclear regions to the extended tidal features. Such kinematic structure cannot survive for
long, so we must be catching IRAS 23128-4250 in a very transient stage associated with the
final merging. Finally, the fourth system (IRAS 20551-4250) consists of a single nucleus in
an r$^{1/4}$ galaxy with a single long tidal tail. The H$\alpha$ is very centrally
concentrated, and shows simple rotational motion indicative of the quiescent dynamical
stage following the completion of the merger. The fact that we see four ultraluminous
systems in four very different dynamical phases argues that 
(at least in this small sample) there is no common dynamical trigger for ULIRG
activity. A similar conclusion was reached by Hibbard \& Yun (1996) from a study of the HI
{\it morphologies} of ULIRGs, which showed no tendency towards prograde interactions.

We are left then with a bit of a dissatisfying -- although perhaps not unexpected -- result.
Both theoretical and observational arguments indicate there is no {\it unique} 
trigger for ULIRG activity. While ULIRGs are associated with
late stage mergers, beyond that there seems to be no one-to-one mapping of dynamics to 
ULIRG activity. Internal structure, orbital dynamics, gas content, and starburst physics
must all play competing and tangled roles in the ultimate triggering of ULIRG activity.

\section{The Believability of N-body Models}

Given the ever-expanding role numerical simulation plays in the study of galactic dynamics,
and in particular galaxy mergers, it is perhaps prudent here to make a few critical comments 
on the robustness of N-body modeling. 
With respect to the ULIRG question, the first obvious shortcoming of the current generation
of N-body models is numerical resolution. The spatial resolution of models such as those
of MH96 or BH96 are $\sim 100$ pc, many orders of magnitude larger than any central accretion 
disk.\footnote{Gasdynamical models with a variable hydrodynamical smoothing length
may purport to have finer resolution in the gas phase, but ultimately the detailed dynamics
cannot be resolved on scales smaller than the gravitational softening length.}
 While these models have shown the efficacy of mergers at driving
radial inflows, they {\it cannot} address accretion onto an AGN, other than the
first step of fueling gas inwards from the disk. This is no trivial matter -- to reach the
accretion disk the gas must shed several orders of magnitude more angular momentum (\eg Phinney
1994). Ideas with which to mediate this further inflow abound, such as nuclear bars (Shlosman
\etal 1990), dynamical friction (Heller \& Schlosman 1994), or gravitational torques (Bekki 1995). While invoking such processes is quite reasonable, we
must realize that in the context of AGN triggering these arguments remain purely speculative,
and cannot be resolved in present models. 

Modeling of the nuclear dynamics of mergers is both a technical and physical challenge. First,
because the resolution scales with the mean interparticle separation ($\langle r \rangle \sim
N^{1/3}$), to get a factor of two improvement in resolution demands an order of magnitude
increase in the number of particles (and CPU time) employed. However, sheer brute force will
not solve the problem -- on these smaller scales the starburst and AGN physics begin to dominate
the dynamical equations. To see this, equate starburst power to binding energy for a 
$M_g = 10^{10}$ M\sun, $10^{12}$ L\sun, $10^7$ year starburst inside 100 pc:
\[
\begin{array}{rcl}
\epsilon L \Delta t & = & GM_g^2/R\\
\epsilon 10^{60} & = & 10^{59}\\
\end{array}
\]
If the efficiency of energy deposition into the ISM ($\epsilon$) is even a few percent,
it can have a significant effect on the nuclear gasdynamics.
Star formation and feedback remains poorly understood, and efforts to incorporate it into
dynamical simulations are fraught with uncertainties -- this problem stands as the biggest
obstacle in modeling the dynamical evolution of ULIRGs. {\it Until better models exist for 
incorporating feedback into N-body simulations, improved spatial resolution is meaningless.}

What, then, {\it can} we believe from N-body simulations? Surely the gravitational dynamics are
well understood, right? They are, up to a point. Unless Newtonian mechanics are wrong, N-body
models accurately calculate the gravitational forces acting on the merging galaxies. The uncertainties
lie not in the physics of gravity, but in the initial conditions of the model, in particular the
mass distribution of the different components of the galaxies. The dynamics of inflow are dependent
on the mass distribution in the inner disk, but are disk galaxies maximal disks? Minimal disks?
Equally problematic is the strong dependency of the merger evolution on the distribution of
dark matter at large radius, which affects the orbital evolution and merging timescale. As
we shall see next, these
uncertainties make even pure gravitational modeling of merging galaxies uncertain.

\section{The Role of Dark Matter}

Dark matter halos play the dominant role in determining the dynamics of merger on large (tens of kpc)
size scales. On these scales, it is the dynamical friction of the dark halos which
brakes the galaxies on their orbit and causes them to merge. Different dark matter halos lead to
different orbital evolution and merging timescales for the colliding galaxies. With the amount of dark matter in
galaxies poorly constrained, particularly at large distances from the luminous disks, these effects
represent a serious uncertainty in dynamical modeling of galaxy mergers. While recent
cosmological simulations give detailed predictions of the dark matter distribution on large scales
(\eg Navarro \etal 1995),
these predictions are at often at odds with observed galaxy rotation curves (McGaugh \& de Blok 1998).

Most models of galaxy mergers to date have typically employed relatively low mass dark halos truncated outside 
of a few tens of kpc. More recently, efforts have been made to include more massive and extended halos
in merger models. While results on the detailed evolution of the tidal debris remain
contentious (Dubinski \etal 1996, 1999; Barnes 1998; Springel \& White 1998), one thing is clear -- the more massive the dark halo is, the longer the merging time.
At first this may seem counter-intuitive, since the more ``braking material'' there is, the faster
the braking ought to be! But halo mass also provides acceleration, so that galaxies with more massive
halos are moving faster at perigalacticon, diminishing the efficiency of dynamical friction. At fixed
circular velocity, the higher encounter velocity wins out over the increased dynamical friction, and
merging time increases. This can be seen in Figure 3, which shows the orbital evolution of two equal-mass
mergers, both with similar rotation curves in the luminous portion of the galaxy, but where one has
a dark matter halo three times the mass of, and twice as extended as, the other.

The differences in orbital evolution among models with different dark halos have several ramifications
for merger dynamics and the formation of ULIRGs in particular; for example:
\begin{itemize}
\item {\sl Timing of inflows:} In \S 2, statistical arguments
were made that the onset of activity is largely suppressed over 80\% of the merger evolution, and that inflow
occurs only late. If in fact halos are even more massive, and the merging timescale even longer,
this constraint becomes even more severe -- over more than 95\% of the merger timescale the galaxies must
lie dormant before activity is triggered. In this case, the dynamical stability must be strong indeed.
\item {\sl Modeling of specific systems:} The rapid advances in N-body modeling
have made it easy to construct 
``made-to-order'' models of specific systems. However, uncertainties in the dark matter distribution
translate in significant uncertainties in the dynamical evolution of mergers inferred from these models.
For example, dynamical models of NGC 7252 can be constructed using a variety
of halo models (Mihos \etal 1998), all of which successfully reproduce the
observed kinematics of the system, yet have orbital characteristics and 
merging timescales which differ significantly.
These uncertainties argue that such specific models are caricatures of the real systems, and that inferences of the {\it detailed} dynamics
of specific systems based on such models are ill-motivated.

\end{itemize}

\begin{figure}
\centerline{\epsfbox{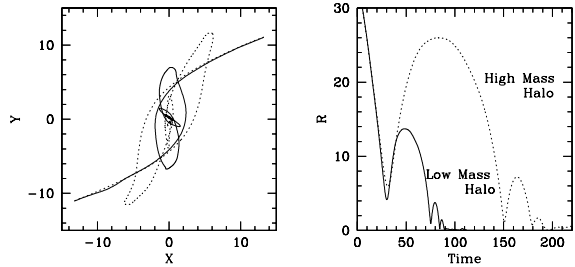}}
\caption{Orbital evolution of a $r_{peri}=4h$ equal-mass collision. Solid line shows the evolution of models
with isothermal halos with mass M=5.8$M_{disk}$ and truncation radius 10$h$, while the dotted
line shows an identical collision between galaxies with halo masses M=17.4$M_{disk}$ and truncation radius 20$h$.}
\end{figure}

\section{High-z Musings}

Finally, in light of the new results from SCUBA on possible high redshift analogs of nearby ULIRGs (\eg
Smail \etal 1998; Barger \etal 1998), it
is interesting to ask how any of the lessons we have learned apply to mergers at higher redshift. The
first immediate question is: are these SCUBA sources disk galaxy mergers at all? The ``smoking gun'' of
a disk merger is the presence of tidal tails, but such features will be difficult to detect.
The surface density of these structures evolves rapidly, giving them a dynamical lifetime which is 
short -- $\sim {\rm\ a\ few\ }t_{rot}(R)$.
Only outside of $\sim$ 10 kpc are tidal features long-lived, yet here their surface brightness is quite faint
($\mu_R$ \gtsima 25--26 mag arcsec$^{-2}$).
Add to this the $(1+z)^4$ cosmological dimming, and by $z=2$ the surface brightness of these features will
be down to 30--31 mag arcsec$^{-2}$, very hard indeed to detect. In fact, with the tidal features so
faint and the inner regions perhaps highly obscured, simply {\it detecting} these galaxies may be quite problematic,
much less determining their structural and dynamical properties.

Nonetheless it is instructive to ask how disk mergers might evolve differently at high redshift compared
to present day mergers. High redshift disks may well have been more gas-rich than current disk galaxies,
resulting in more fuel for star formation and in a lower disk stability threshold (the Toomre $Q$). As a
result, rather than driving gas inwards, collisions of galaxies at high redshift may instead result
in pockets of disk gas going into local collapse and increasing disk star formation 
at the expense of nuclear starbursts. Morphologically, we might expect mergers to show an
extended, knotty structure of sub-luminous clumps rather than an extremely luminous, nucleated structure.
Qualitatively this is similar to the types of objects found in the Hubble Deep Field(s), where multiple
bright knots observed in the restframe UV may be embedded in a single structure when observed
in rest frame optical. We should therefore apply extreme caution when applying results obtained
from merger models based on nearby galaxies to the high redshift universe.

\end{document}